\documentclass{birkjour}
 \newtheorem{thm}{Theorem}[section]

 \theoremstyle{definition}
 
 \theoremstyle{remark}
 \newtheorem{rem}[thm]{Remark}
 
 \numberwithin{equation}{section}

  \newcommand{\nn}{\nonumber}
  
  \newcommand{\al}{\alpha}

  \newcommand{\Ga}{\Gamma}

  \newcommand{\eps}{\varepsilon}
  \newcommand{\ka}{\kappa}

  \newcommand{\la}{\lambda}
  \newcommand{\La}{\Lambda}
  \newcommand{\si}{\sigma}
  \newcommand{\vf}{\varphi}
  \newcommand{\om}{\omega}

   \newcommand{\ut}{\tilde u}
  \renewcommand{\d}{\partial}
  \newcommand{\be}{\begin{equation}}
  \newcommand{\ee}{\end{equation}}
  \newcommand{\NN}{\mathbb{N}} % Insieme numeri naturali
  \newcommand{\RR}{\mathbb{R}} % Insieme numeri reali
  \newcommand{\CC}{\mathbb{C}} % insieme numeri complessi

  \newcommand{\Dcal}{\mathcal{D}}

  \newcommand{\Hcal}{\mathcal{H}}

\begin{document}

%-------------------------------------------------------------------------
% editorial commands: to be inserted by the editorial office
%
%\firstpage{1} \volume{228} \Copyrightyear{2004} \DOI{003-0001}
%\seriesextraline{This is the Concrete Title of this Book\br H.E. R and S.T.C. W, Eds.}
%\firstpage{1}

\title[Higgs Mechanism and RG-Flow]
 {Higgs Mechanism and Renormalization Group Flow:  Are They Compatible?}

%----------Author 1
\author[M.~D\"utsch]{Michael D\"utsch}

\address{%
Institut f\"ur Theoretische Physik\\
 Universit\"at G\"ottingen\\
Friedrich-Hund-Platz 1\\ 
D-37077 G\"ottingen\\
Germany}

\email{michael.duetsch@theorie.physik.uni-goe.de}

%\thanks{This work was completed with the support of our \TeX-pert.}


%\date{January 1, 2004}
%\dedicatory{}
%%% ----------------------------------------------------------------------

\begin{abstract}
Usually the Lagrangian of a model for massive vector bosons is derived in a geometric 
way by the Higgs mechanism. We investigate whether this geometric structure is maintained 
under the renormalization group (RG) flow. Using the framework of Epstein-Glaser 
renormalization, we find that the answer is 'no', if the 
renormalization mass scale(s) are chosen in a way corresponding 
to the minimal subtraction scheme. This result is derived 
for the $U(1)$-Higgs model to 1-loop order. On the other
hand we give a model-independent proof that 
physical consistency, which is a weak form of BRST-invariance
of the time-ordered products, is stable under the RG-flow.
\end{abstract}

%%% ----------------------------------------------------------------------
\maketitle
%%% ----------------------------------------------------------------------
%\tableofcontents
\section{Introduction}

By the renormalization group (RG) flow we have a tool to describe a QFT-model
at different scales. In this description, the basic fields, the gauge-fixing parameter,
the masses and the prefactors of the various interaction terms are scale-dependent 
quantities.

On the other hand the derivation of the Lagrangian of a
model for massive vector bosons by the Higgs mechanism, i.e.~by spontaneous 
symmetry breaking of a gauge theory, implies that the prefactors 
of the various interaction terms are uniquely determined functions of the coupling 
constant(s) and masses.

Do these functions remain unchanged under the RG-flow, i.e.~under an arbitrary change of 
scale? This question is a reformulation of the title of this paper. Since the 
non-trivial contributions to the RG-flow come from loop diagrams and 
different interaction terms get different 
loop-corrections, it is uncertain, whether the answer is 'yes'.
Or - one can come to the same conclusion by considering the underlying frameworks:
the Higgs mechanism is formulated in classical field theory and, to the best of our 
knowledge, it is not understood in a pure QFT framework; on the other hand, 
the RG-flow is a pure quantum effect.

Some readers may wonder, whether the Lagrangian of the scaled model describes still 
a consistent QFT-model, if it is not derivable by the Higgs mechanism? The 
answer is 'yes', for the following reasons: since  Poincar{\'e} invariance, 
relevant discrete symmetries and renormalizability are maintained under the RG-flow,
the crucial requirement for consistency of a quantum gauge model is physical consistency (PC) 
\cite{Kugo-Ojima,DS2000}. This is the condition that the free BRST-charge%
\footnote{That is the charge implementing the BRST-transformation of the asymptotic free fields.}
commutes with the ``$S$-matrix'' in the adiabatic limit, in order that the latter induces a 
well-defined operator on the physical subspace. We give a model-independent proof that PC 
is maintained under the RG-flow (Theorem \ref{thm:PC}).

In the literature we could not find an explicit 'yes' or 'no' to the question
in the title. However, 
some papers silently assume that the answer is 'yes' -- see the few examples mentioned in
\cite[Introduction]{Duetsch15}. The answer certainly depends on the renormalization scheme.

We work with the definition of the RG-flow given in framework
Epstein-Glaser renormalization \cite{EG73}: since a scaling transformation 
amounts to a change of the renormalization prescription, it can equivalently be expressed
by a renormalization of the interaction -- this is an application of the
Main Theorem, see \cite{PS82,HW03,DF04,BDF09}. The so defined RG-flow depends on the 
renormalization scheme via the two possibilities that the scaling transformation may act 
on the renormalization mass scale(s) or it may not; and this may be different for
different Feynman diagrams. 

We investigate the question in the title by explicit 1-loop calculations -- the technical 
details are omitted in this paper, they are given in \cite{Duetsch15}. To minimise the
computations, we study the model of one massive vector field, that is, we start the RG-flow 
with the $U(1)$-Higgs model. 

\section{Precise formulation of the question}

\paragraph{Lagrangian of the initial model}
The just mentioned model has one massive vector field $A^\mu$, the 
  corresponding St\"uckelberg field $B$, a further real scalar field $\vf$ 
(``Higgs field'') and the  Fadeev-Popov ghost fields $(u\,,\,\tilde u)$.
The Lagrangian reads
\be\label{eq:L-total}
  L_\mathrm{total}\simeq
  -\frac{1}4\,F^2+\frac{1}2\,(D^\mu\Phi)^*D_\mu\Phi-V(\Phi)+L_\mathrm{gf}+L_\mathrm{ghost}\ ,
  \ee
  where $\simeq$ means equal up to the addition of terms of type $\d^a A$, where
$|a|\geq 1$ and $A$ is a local field polynomial. In addition
we use the notations
 $F^2:=(\d^\mu A^\nu-\d^\nu A^\mu)(\d_\mu A_\nu-\d_\nu A_\mu)$,
  \be\label{geom1}
  \Phi:=iB+\frac{m}{\ka}+\vf\ ,\quad D^\mu:=\d^\mu-i\ka\,A^\mu
  \ee
  and 
  \be\label{geom2}
  V(\Phi):=\frac{\ka^2m_H^2}{8m^2}\,(\Phi^*\Phi)^2-\frac{m_H^2}{4}\,(\Phi^*\Phi)+\frac{m_H^2 m^2}{8\ka^2}\ ,
  \ee
where $\kappa$ is the coupling constant and $m$ and $m_H$ are the masses of the $A$- and $\vf$-field,
respectively, as it turns out below in \eqref{L-free}.
The gauge-fixing and ghost Lagrangian are given by
\be\label{L-gf}
  L_\mathrm{gf}:=-\frac{\La}{2} \biggl( \d A + \frac{m}{\La} B \biggr)^2
  \ee  
and
\be\label{L-ghost}
  L_{\mathrm{ghost}}:= \d\ut\d u - \frac{m^2}{\La}\, \ut u
   - \frac{\ka\,m}{\La}\,\ut u\vf\ ,
  \ee 
respectively, where $\La$ is the gauge-fixing parameter. 
The masses of the $A$- and $\vf$-field are generated by the Higgs mechanism.

In view of perturbation theory we
split $L_\mathrm{total}$ into a free part $L_0$ (all bilinear terms)
and an interacting part $L$ (all tri- and quadrilinear terms):
\begin{align}
 L_0 &= -\frac{1}4\,F^2 + \frac{m^2}{2} A^2 
  + \frac{1}{2} (\d B)^2 -\frac{m^2}{2\La}\, B^2 
  - \frac{\La}{2} (\d A)^2 \nn\\ 
  &+ \frac{1}{2} (\d\vf)^2 - \frac{m_H^2}{2}\, \vf^2 
  + \d\ut\d u - \frac{m^2}{\La}\, \ut u \ , \label{L-free}\\
 L&= \ka\Bigl( m\, A^2\vf - \frac{m^2}{\La}\ut u\vf + 
B(A\d\vf) - \vf(A\d B) - \frac{m_H^2}{2m} \vf^3
  - \frac{m_H^2}{2m} B^2\vf\Bigr)\nn \\
  &+ \ka^2\Bigl(\frac{1}{2} A^2 (\vf^2 + B^2)
  - \frac{m_H^2}{8m^2} \vf^4 - \frac{m_H^2}{4m^2} \vf^2 B^2
  - \frac{m_H^2}{8m^2} B^4\Bigr)\ , \label{L-int}
\end{align}
where $V^2:=V^\mu V_\mu$, $VW:=V^\mu W_\mu$ for Lorentz vectors $V,W\in\CC ^4$.

\begin{rem}
The BRST-transformation is a graded derivation which commutes with partial derivatives and 
  is given on the basic fields by
  \begin{align}\label{BRS-trafo}
  s\,A^\mu=\d^\mu u\ ,& \quad s\,B=mu+\ka\,u\vf\ ,\quad s\,\vf=-\ka\,Bu\ ,\nn\\
  s\,u=0\ ,& \quad s\,\ut=-\La\,(\d A+\frac{m}{\La}\,B)\ .
  \end{align}
By $s_0:=s\vert_{\ka =0}$ we denote its version
for the free theory. We point out that $L$ and $L_0$ are 
invariant w.r.t.~the pertinent BRST-transformation:
\be
sL\simeq 0\ ,\quad\quad s_0L_0\simeq 0\ ,
\ee
where $\simeq$ has the same meaning as above.
\end{rem}

\paragraph{Definition of the RG-flow}
In view of Epstein-Glaser renormalization \cite{EG73} we write
\be\label{L1+L2}
L=\ka\,L_1+\ka^2\,L_2
\ee
and introduce an adiabatic switching of 
the coupling constant by a test function $g\in\mathcal{D}(\RR ^4)$:
\be\label{L(g)}
L(g)\equiv L^{\mathbf{m}}(g):=\int dx\,\Bigl(\ka\,g(x)\,L_1(x)+
\bigl(\ka\,g(x)\bigr)^2\,L_2(x)\Bigr)\ .
\ee
For later purpose we have introduced the upper index $\mathbf{m}:=(m,m_H)$.

In the Epstein-Glaser framework the RG-flow is defined 
by a scaling transformation $\sigma_\rho$ \cite{HW03,DF04,BDF09}: 
  \be\label{scaltrafo}
  \si_\rho^{-1}(\phi(x))=\rho\,\phi(\rho x)\ ,\quad \phi=A^\mu,B,\vf,u,\ut\ ,\quad\rho > 0\ ,
  \ee
  and a simultaneous scaling of the masses 
  $\mathbf{m}\mapsto\rho^{-1}\mathbf{m}=(\rho^{-1} m,\rho^{-1} m_H)$;
see \cite{DF04} for the precise definition of $\si_\rho$.
Under this transformation the classical action is invariant
(up to a scaling of the switching function $g$).

  In QFT scaling invariance is in general broken in the process of renormalization. To explain this, 
we introduce the generating functional $S(iL(g))$ of the 
  time-ordered products of $L(g)$, i.e.
\be\label{T-product}
T_n(L(g)^{\otimes n})=\frac{d^n}{i^n\,d\eta^n}\vert_{\eta =0}\,S(i\eta \,L(g))
\quad\text{or generally}\quad T_n=S^{(n)}(0)\ ,
\ee
which we construct inductively by Epstein-Glaser renormalization \cite{EG73}.
To define the RG-flow, we need to perform the adiabatic limit 
  \be\label{adlim}
  \mathbf{S}[L]:=\lim_{\eps\downarrow 0}S(iL(g_\eps))\ ,
\quad g_\eps(x):=g(\eps x)\ ,
  \ee
where $g(0)=1$ is assumed.
For a purely massive model and with a suitable (re)normalization of $S(iL(g))$,
this limit exists in the strong operator sense. For a rigorous proof of this statement we refer
to \cite{EG73,EG76}; in this paper we treat the adiabatic limit on a heuristic level.

The Main Theorem of perturbative  renormalization \cite{PS82,DF04,HW03}
   implies that a scaling transformation of $\mathbf{S}[L]$, i.e.
$$
\mathbf{S}_{\mathbf{m}}[L^{\mathbf{m}}]\mapsto 
\sigma_\rho(\mathbf{S}_{\rho^{-1}\mathbf{m}}[\si_\rho^{-1}(L^{\mathbf{m}})])\ ,
$$
  can equivalently be expressed by a renormalization of the interaction $L^{\mathbf{m}}
\mapsto z_\rho(L^{\mathbf{m}})$ (``running interaction''), explicitly
  \be\label{main-theorem-adlim}
  \sigma_\rho(\mathbf{S}_{\rho^{-1}\mathbf{m}}[\si_\rho^{-1}(L^{\mathbf{m}})])=
\mathbf{S}_{\mathbf{m}}[ z_\rho(L^{\mathbf{m}})]\ ,
  \ee
where the lower index $\mathbf{m}$ of $\mathbf{S}_{\mathbf{m}}$ denotes the masses of the Feynman propagators.
  This is explained in detail in Sect.~\ref{sec:stab-PC}. 

\paragraph{The form of the running interaction}
Using general properties of the running interaction (derived in \cite{DF04}), 
we know that  each term appearing in $z_\rho(L)$ is Lorentz invariant, has ghost number $=0$ and
  has mass dimension $\leq 4$. In addition, using that $L$ \eqref{L-int}
is even under the field parity transformation
  \be\label{fieldparity}
  (A,B,\vf,u,\ut)\mapsto (-A,-B,\vf,u,\ut)\ ,
  \ee 
one easily derives that also $z_\rho(L)$ is even under this transformation. 
One can also show that only one term containing the Fadeev-Popov ghosts can appear in
 $(z_\rho(L)-L)$, namely a term $\sim\ut u$. Moreover,
with a slight restriction on the (re)normalization of $S(iL(g))$, one can exclude 
$1$-leg terms from $z_\rho(L)$ \cite{Duetsch15}. 

Using these facts, we conclude that the running interaction has the form
  \begin{align}\label{z(L)}
   z_\rho(L)  \simeq &\hbar^{-1}\Bigl[ k_\rho   
   -\frac{1}4 a_{0\rho}\,F^2 + \frac{m^2}{2}a_{1\rho} A^2 - \frac{a_{2\rho}}{2} (\d A)^2
  + \frac{1}{2} b_{0\rho} (\d B)^2 - \frac{m^2}{2\,\La}\, b_{1\rho}\, B^2\nn\\ 
  &+ \frac{1}{2} c_{0\rho}\,(\d\vf)^2 - \frac{m_H^2}{2}c_{1\rho}\, \vf^2 
   -\frac{m^2}{\La} \,c_{2\rho}\, \ut u+b_{2\rho}\,m\,(A\d B)\nn\\ 
  &+\ka\Bigl((1+ l_{0\rho})\,m\, A^2\vf -\frac{m}{\La}\,\ut u\vf +(1+ l_{1\rho})\, B(A\d\vf)\nn\\
 & - (1+ l_{2\rho})\, \vf(A\d B) - \frac{(1+ l_{3\rho}) m_H^2}{2m}\, \vf^3
  - \frac{(1+l_{4\rho}) m_H^2}{2m}\, B^2\vf\Bigr)\nn \\
  &+\ka^2\Bigl( \frac{(1+ l_{5\rho})}{2} A^2\, \vf^2 + \frac{(1+ l_{6\rho})}{2} A^2\, B^2
  - \frac{(1+ l_{7\rho})m_H^2}{8m^2}\, \vf^4 \nn\\
&- \frac{(1+ l_{8\rho})m_H^2}{4m^2}\, \vf^2 B^2
  - \frac{(1+ l_{9\rho})m_H^2}{8m^2}\, B^4 +l_{11\rho}\,(A^2)^2\Bigr)\Bigr]\ ,
  \end{align}
  where $\simeq$ has the same meaning as in \eqref{eq:L-total} and
$k_\rho\in\hbar\,\CC[[\hbar]]$ is a constant field (it is the contribution of the
vacuum diagrams), which may be neglected.

The dimensionless, $\rho$-dependent coefficients
  $k_\rho,\,a_{j\rho},\,b_{j\rho},\,c_{j\rho}$ and $l_{j\rho}$ 
will collectively be denoted by $e_\rho$. In principle these coefficients are computable -- 
at least to lowest orders; however, at the present stage they are
unknown. As shown in \cite{Duetsch15}, the $e_\rho$'s are formal power series in $\ka^2\hbar$ with 
  vanishing term of zeroth order,
  \be\label{powerseries}
  e_\rho=\sum_{n=1}^\infty e_\rho^{(n)}\,(\ka^2\hbar)^n\ ,\quad e=k,a_j,\,b_j,\,c_j,\,l_j\ .
  \ee
  Due to $z_{\rho=1}(L)=L/\hbar$, all
functions $\rho\mapsto e_\rho$ have the initial value $0$ at $\rho=1$. 

\paragraph{Renormalization of the wave functions, masses, gauge-fixing parameter and coupling parameters}
  Except for the $A\d B$- and $A^4$-term, all field monomials appearing in $z_\rho(L)$ are already present in
$L_0+L$. Therefore, introducing new fields, which are of the form
\be\label{new-fields}
\phi_\rho(x)=f_\phi(\rho)\,\phi(x)\ ,\quad \phi=A,\,B,\,\vf,
\ee
where $f_\phi:\,(0,\infty)\to\CC$ is a $\phi$-dependent function,
and introducing a running gauge-fixing parameter $\La_\rho$,
  running masses $\mathbf{m}_\rho\equiv (m_\rho,\,m_{B\rho},\,m_{u\rho},\,m_{H\rho})$ and 
  running coupling constants $\ka_\rho\la_{j\rho}$, we can 
achieve that $L_0+z_\rho(L)-k_\rho$ has roughly the same form as $L_0+L$: 
\be\label{parameter-ren}
L_0+z_\rho(L)-k_\rho=L_0^\rho+L^\rho\ ,
\ee
where
\begin{align}
 L^\rho_0 &= -\frac{1}4\,F_\rho^2 + \frac{m_\rho^2}{2} A_\rho^2 
  + \frac{1}{2} (\d B_\rho)^2 -\frac{m_{B\rho}^2}{2}\, B_\rho^2 
  - \frac{\La_\rho}{2} (\d A_\rho)^2 \nn\\ 
  &+ \frac{1}{2} (\d\vf_\rho)^2 - \frac{m_{H\rho}^2}{2}\, \vf_\rho^2 
  + \d\ut\d u - m_{u\rho}^2\, \ut u \ , 
\end{align}
(with $F_\rho^{\mu\nu}:=\d^\mu A_\rho^\nu-\d^\nu A_\rho^\mu$) and
\begin{align}
L^\rho &=\ka_\rho\Bigl( m_\rho A_\rho^2\vf_\rho - \frac{\la_{10\rho}\,m_{u\rho}^2}{m_\rho}\ut u\vf + 
\la_{1\rho} B_\rho(A_\rho\d\vf_\rho)\nn\\ 
  &- \la_{2\rho} \vf_\rho(A_\rho\d B_\rho) - \frac{\la_{3\rho} m_{H\rho}^2}{2m_\rho} \vf_\rho^3
  - \frac{\la_{4\rho} m_{H\rho}^2}{2m_\rho} B_\rho^2\vf_\rho\Bigr)\nn \\
  &+ \ka^2\Bigl(\frac{\la_{5\rho}}{2} A_\rho^2 \vf_\rho^2 + \frac{\la_{6\rho}}{2} A_\rho^2 B_\rho^2
  - \frac{\la_{7\rho}m_{H\rho}^2}{8m_\rho^2} \vf_\rho^4 \nn\\
  &- \frac{\la_{8\rho}m_{H\rho}^2}{4m_\rho^2} \vf_\rho^2 B_\rho^2
  - \frac{\la_{9\rho}m_{H\rho}^2}{8m_\rho^2} B_\rho^4+\la_{11\rho}\,A_\rho^2\Bigr)\nn\\
  &+\bigl((\la_{12\rho}-1)m_\rho+\sqrt{\La_\rho}\,m_{B\rho}\bigr)\,A_\rho\d B_\rho\ .
  \end{align}
In view of the Higgs mechanism for $L+z_\rho(L)$ \eqref{L-new}, the definition of $\la_{12\rho}$ is 
rather complicated. Apart from the $A\d B$-term, we have absorbed the novel 
bilinear interaction terms in the free Lagrangian.
Since every new field is of the form \eqref{new-fields},
the condition \eqref{parameter-ren} is an equation for polynomials in the old fields; 
equating the coefficients we obtain the following explicit formulas for the running quantities:\\
  - for the wave functions
  \be\label{wf-ren}
  A^\mu_\rho=\sqrt{1+a_{0\rho}}\,A^\mu\ ,\quad
  B_\rho=\sqrt{1+b_{0\rho}}\,B\ ,\quad
  \vf_\rho=\sqrt{1+c_{0\rho}}\,\vf\ ;
  \ee
  - for the gauge-fixing parameter
  \be\label{gauge-ren}
  \La_\rho=\frac{\La+a_{2\rho}}{1+a_{0\rho}}\ ;
  \ee
  - for the masses
  \begin{align}\label{mass-ren}
  m_\rho=\sqrt{\frac{1+a_{1\rho}}{1+a_{0\rho}}}\,m\ ,&\qquad
  m_{H\rho}=\sqrt{\frac{1+c_{1\rho}}{{1+c_{0\rho}}}}\,m_H\ ,\nn\\
  m_{B\rho}=\sqrt{\frac{1+b_{1\rho}}{{1+b_{0\rho}}}}\,\frac{m}{\sqrt{\La}}\ ,&\qquad
  m_{u\rho}=\sqrt{1+c_{2\rho}}\,\frac{m}{\sqrt{\La}}\ ;
  \end{align}
  - for the coupling constant
  \be\label{cc-ren}
  \kappa_\rho=\frac{1+l_{0\rho}}{\sqrt{(1+a_{0\rho})(1+a_{1\rho})(1+c_{0\rho})}}\,\kappa\ ; 
  \ee
  and the running coupling parameters $\la_{j\rho}$ are determined analogously. 

By the renormalization of the wave functions, masses and gauge 
fixing-parameter, we change the splitting of the total Lagrangian $L_0+z_\rho(L)$ into a free and 
interacting part, i.e.~we change the starting point for the perturbative expansion. To justify this,
one has to show that the two pertubative QFTs given by the splittings $L_0+z_\rho(L)$ and
$L_0^\rho+L^\rho$, respectively, have the same physical content.\footnote{This statement can be viewed as an 
application of the ``Principle of Perturbative Agreement'' of Hollands and Wald \cite{HW05}.}
Using the framework of algebraic QFT, one has to show the following: given a renormalization
prescription for $L_0+z_\rho(L)$, there exists a  renormalization prescription for $L_0^\rho+L^\rho$, 
such that, in the 
algebraic adiabatic limit, the pertinent nets of local observables (see \cite{BF00} or \cite{DF04,BDF09}) 
are equivalent. This task is beyond the scope of this paper. 

\paragraph{Higgs mechanism at an arbitrary scale}
Our main question is whether the Lagrangian $L_0^\rho+L^\rho$ can also be 
derived by the Higgs mechanism for all $\rho >0$. By the latter we mean 
  \be\label{L-new}
  L_0^\rho+L^\rho\simeq
-\frac{1}4\,F_\rho^2+\frac{1}2\,(D_\rho^\mu\Phi_\rho)^*D_{\rho\mu}\Phi_\rho-V_\rho(\Phi_\rho)
 + L_{\mathrm{gf}}^\rho+L_{\mathrm{ghost}}^\rho\ ,
  \ee
  where $\Phi_\rho,\,D_\rho$ and $V_\rho(\Phi_\rho)$ are obtained from \eqref{geom1}-\eqref{geom2}
by replacing \newline
$(A^\mu,\,B,\,\vf,\,m,\,m_H,\,\La)$ by 
$(A^\mu_\rho,\,B_\rho,\,\vf_\rho,\,m_\rho,\,m_{H\rho},\,\La_\rho)$
  and
  \begin{align}\label{L-gf-ghost-rho}
  L_{\mathrm{gf}}^\rho & :=-\frac{\La_\rho}{2} \biggl( \d A_\rho +
  \frac{m_{B\rho}}{\sqrt{\La_\rho}} B_\rho \biggr)^2\ ,\nn\\
  L_{\mathrm{ghost}}^\rho& :=\d\ut\cdot \d u-m_{u\rho}^2\,\ut u
  -\frac{\ka_\rho\,\la_{10\rho}\,m_{u\rho}^2}{m_\rho}\,\ut u\vf_\rho\ .
  \end{align}
For the property \eqref{L-new} we also say that the model ``can be geometrically interpreted as a spontaneously
broken gauge theory at all scales'' \cite{Duetsch15}. By a straightforward calculation we find that \eqref{L-new}
 is equivalent to
  \be\label{runningparameter-geom}
  \la_{1\rho}=\la_{2\rho}=...=\la_{9\rho}=1\ ,\quad\la_{11\rho}=\la_{12\rho}=0\ .
  \ee

To simplify the calculations we assume that initially we are in Feynman gauge: $\La_{\rho=1}=1$.
With that the geometrical interpretability \eqref{runningparameter-geom} 
is equivalent to the following relations among the coefficients $e_\rho$:
\begin{align}
\la_{1\rho}=1\,\,\,\text{gives}&\quad\quad\quad\frac{1+l_{1\rho}}{1+l_{0\rho}}=
\sqrt{\frac{1+b_{0\rho}}{1+a_{1\rho}}}\ ,\label{e-geom-interpret1}\\
\la_{2\rho}=1\,\,\,\text{gives}&\quad\quad\quad l_{2\rho}=l_{1\rho}\ ,\label{e-geom-interpret2}\\
\la_{3\rho}=1\,\,\,\text{gives}&\quad\quad\quad\frac{1+l_{3\rho}}{1+l_{0\rho}}=
\frac{1+c_{1\rho}}{1+a_{1\rho}}\ ,\label{e-geom-interpret3}\\
\la_{4\rho}=1\,\,\,\text{gives}&\quad\quad\quad\frac{1+l_{4\rho}}{1+l_{3\rho}}=
\frac{1+b_{0\rho}}{1+c_{0\rho}}\ ,\label{e-geom-interpret4}\\
%\end{align}
%\begin{align}
\la_{5\rho}=1\,\,\,\text{gives}&\quad\quad\quad\frac{1+l_{5\rho}}{(1+l_{0\rho})^2}=
\frac{1}{1+a_{1\rho}}\ ,\label{e-geom-interpret5}\\
\la_{6\rho}=1\,\,\,\text{gives}&\quad\quad\quad\frac{1+l_{6\rho}}{1+l_{5\rho}}=
\frac{1+b_{0\rho}}{1+c_{0\rho}}\ ,\label{e-geom-interpret6}
\end{align}
\begin{align}
\la_{7\rho}=1\,\,\,\text{gives}&\quad\quad\quad\frac{1+l_{7\rho}}{(1+l_{0\rho})^2}=
\frac{1+c_{1\rho}}{(1+a_{1\rho})^2}\ ,\label{e-geom-interpret7}\\
\la_{8\rho}=1\,\,\,\text{gives}&\quad\quad\quad\frac{1+l_{8\rho}}{1+l_{7\rho}}=
\frac{1+b_{0\rho}}{1+c_{0\rho}}\ ,\label{e-geom-interpret8}\\
\la_{9\rho}=1\,\,\,\text{gives}&\quad\quad\quad\frac{1+l_{9\rho}}{1+l_{7\rho}}=
\Bigl(\frac{1+b_{0\rho}}{1+c_{0\rho}}\Bigr)^2\ ,\label{e-geom-interpret9}\\
\la_{11\rho}=0\,\,\,\text{gives}&\quad\quad\quad l_{11\rho}=0\ ,\label{e-geom-interpret11}\\
\la_{12\rho}=0\,\,\,\text{gives}&\quad\quad\quad b_{2\rho}=
\sqrt{(1+a_{2\rho})(1+b_{1\rho})}-\sqrt{(1+a_{1\rho})(1+b_{0\rho})}\ .\label{e-geom-interpret12}
\end{align}
%Searching all values for the coefficients $e_\rho$ which solve
%this system of equations, we find that this is quite a large set: neglecting $k_\rho$, 
%$9$ coefficients can freely be chosen (e.g.~$a_{0\rho},\,a_{1\rho},\,a_{2\rho},\,b_{0\rho},\,
%b_{1\rho},\,c_{0\rho},\,c_{1\rho},\,c_{2\rho}$ and $l_{0\rho}$), the other $11$ 
%coefficients are then uniquely determined by the
%$11$ equations \eqref{e-geom-interpret1}-\eqref{e-geom-interpret12}.

Combining the equations \eqref{e-geom-interpret3}, \eqref{e-geom-interpret5} and \eqref{e-geom-interpret7}
we obtain
\be\label{geom-interpret-crucial}
\frac{1+l_{7\rho}}{1+l_{3\rho}}=\frac{1+l_{5\rho}}{1+l_{0\rho}}\ .
\ee
This condition and \eqref{e-geom-interpret12} are crucial for the geometrical interpretability, as 
we will see.

\begin{rem}
BRST-invariance of $L_0+z_\rho(L)$ is a clearly stronger property than the geometrical interpretability
\eqref{L-new}. More precisely: considering the coefficients $e_\rho$ as unknown and assuming that
$s(L_0+z_\rho(L))\simeq 0$, we obtain rather restrictive relations among the coefficients $e_\rho$ which
imply the equations \eqref{e-geom-interpret1}-\eqref{e-geom-interpret12}. Ignoring $k_\rho$, the number of
coefficients $e_\rho$, which are left freely chosable by the BRST-property, is $3$; and for the 
geometrical interpretability this number is $9$ -- see \cite{Duetsch15}.
\end{rem}

\section{Physical consistency and perturbative gauge invariance}
\label{sec:stab-PC}
\paragraph{Physical consistency (PC)}
The generic problem of a model containing
spin $1$ fields, is the presence of unphysical fields. A
way to solve this problem in a scattering framework is to construct $S(iL(g))$
such that the following holds. For the asymptotic free fields let $\Hcal_{\mathrm{phys}}$
be the ``subspace'' of physical states. In the adiabatic limit lim $g\to 1$,
$S(iL(g))$ has to induce  a well defined operator from $\Hcal_{\mathrm{phys}}$ into itself,
which is the physically relevant $S$-matrix.

To formulate this condition explicitly, let $Q$ be the generator of 
the free BRST-transformation $s_0:=s|_{\ka =0}$:
  \be
  [Q,\phi]^\mp_\star\approx\,i\hbar\, s_0\phi\ ,\quad\phi=A^\mu\,,\,B\,,\,\vf\,,\,u\,,\,\ut\ ,
  \ee
where $[\cdot\,,\,\cdot]^\mp_\star$ denotes the graded commutator w.r.t.~the $\star$-product
and $\approx$ means 'equal modulo the free field equations'. With that we may write
 $\Hcal_{\mathrm{phys}}:=\frac{\mathrm{ker}\,Q}{\mathrm{ran}\,Q}$, and the mentioned, fundamental condition
on $S(iL(g))$ is equivalent to
\be\label{PC}
0\approx [Q,\mathbf{S}[L]]_\star\vert_{\mathrm{ker}\,Q}\equiv
\lim_{\eps\downarrow 0}[Q,S(iL(g_\eps)/\hbar)]_\star\vert_{\mathrm{ker}\,Q}\ ,
\ee
see \cite{Kugo-Ojima,DS2000}. For simplicity we omit the restriction 
to $\mathrm{ker}\,Q$ and call the resulting condition ``physical consistency (PC)''.

\paragraph{Stability of PC under the RG-flow}
A main, model-independent result of this paper is that PC is maintained under the RG-flow.
\begin{thm}\label{thm:PC}
Assume that $S_\mathbf{m}(iL(g))$ is renormalized such that the adiabatic limit
$\eps\downarrow 0$ exists and is unique for 
  $\si_\rho\circ S_{\rho^{-1}\mathbf{m}}\circ\si_\rho^{-1}(iL(g_\eps))\ $  $\forall\rho > 0$,
and such that  $S_\mathbf{m}(iL^\mathbf{m}(g))$ fulfills PC 
for all values $m_j>0$ of the masses $\mathbf{m}=(m_j)$. Then, the following holds:
  \be\label{PC-RG}
\bigl[Q,\mathbf{S}[z_\rho(L)]\bigr]_\star\equiv
\lim_{\eps\downarrow 0}[Q,S(iz_\rho(L)(g_\eps)]_\star\approx 0\ ,\quad\forall\rho >0\ .
  \ee
\end{thm}
Hence, at least in this weak form, BRST-invariance of the time-ordered products is stable under the RG-flow.
%And, since for the $U(1)$-Higgs model it is well known that PC can be fulfilled for all values of $m,m_H>0$, 
%the Theorem states that this model is consistent at all scales.

\begin{proof} As a preparation we explain the construction 
of $z_\rho(L)$ and derive \eqref{main-theorem-adlim}. Assuming that
$S$ fulfills the axioms of Epstein-Glaser renormalization, this holds 
also for the scaled time-ordered products $\si_\rho\circ S\circ\si_\rho^{-1}$; therefore, 
the Main Theorem \cite{DF04,HW03} applies: there exists 
a unique map $Z_\rho\equiv Z_{\rho,\mathbf{m}}$ from the space of local interactions into itself such that
  \be\label{main-theorem}
  \si_\rho\circ S_{\rho^{-1}\mathbf{m}}\circ\si_\rho^{-1}=S_{\mathbf{m}}\circ Z_{\rho,\mathbf{m}}
  \ee
  (the lower index $\mathbf{m}$ on $S$ and $Z_\rho$  denotes the masses of the underlying $\star$-product,
  i.e.~the masses of the Feynman propagators). 

In view of the adiabatic limit we investigate $Z_\rho(iL(g_\eps)/\hbar)$
and take into account that $\d g_\eps(x)\sim \eps$.
From \cite[Prop.~4.3]{DF04} we know that there exist local field polynomials 
$p_{k\rho}(L)$ such that
\be
Z_\rho(iL(g_\eps)/\hbar)=\frac{i}{\hbar}\Bigl(L(g_\eps)+\sum_{k=2}^\infty\int dx\,\,
p_{k\rho}(L)(x)\,(\ka g_\eps(x))^k\Bigr)+\mathcal{O}(\eps)\ .
\ee
Obviously, $p_{k\rho}(L)$ is not uniquely determined: one may add
terms of type $\d^a A$, $|a|\geq 1$, where $A$ is a local field polynomial.
Setting
\be
z_\rho(L)(g):=\frac{1}{\hbar}\sum_{k=1}^\infty\int dx\,\Bigl(L_k(x)+
p_{k\rho}(L)(x)\Bigr)\,(\ka g(x))^k\ ,
\ee
where $p_{1\rho}:=0$ and $L_k:=0$ for $k\geq 3$, we obtain
\be\label{Z(L)-adlim}
Z_\rho(iL(g_\eps)/\hbar)=i\,z_\rho(L)(g_\eps)+\mathcal{O}(\eps)\ .
\ee
Using this result and (multi-)linearity of the time-ordered products, 
we obtain \eqref{main-theorem-adlim}:
  \begin{align}\label{MThmadlim}
  \sigma_\rho(&\mathbf{S}_{\rho^{-1}\mathbf{m}}[\si_\rho^{-1}(L^{\mathbf{m}})]):=
\lim_{\eps\downarrow 0}\si_\rho\circ S_{\rho^{-1}\mathbf{m}}\circ\si_\rho^{-1}(iL^{\mathbf{m}}(g_\eps))\nn\\
 & =\lim_{\eps\downarrow 0}S_{\mathbf{m}}\bigl(Z_\rho(iL^{\mathbf{m}}(g_\eps))\bigl)
   =\lim_{\eps\downarrow 0}S_{\mathbf{m}}\bigl(i\,z_\rho(L^{\mathbf{m}})(g_\eps)\bigl)=:
\mathbf{S}_{\mathbf{m}}[ z_\rho(L^{\mathbf{m}})]\ .
  \end{align}
By assumption the limit exists on the l.h.s.; hence, it exists also on the r.h.s..

With these tools we are able to prove \eqref{PC-RG}:
 using the relations
  \be
  \si_\rho^{-1}(L^\mathbf{m}(g))=L^{\rho^{-1}\mathbf{m}}(g_{1/\rho})\,\,\, 
\quad (\,\text{again}\,\,g_\la(x):=g(\la x)\,)
  \ee
  and
  \be 
  \si_\rho(F\star_{\rho^{-1}\mathbf{m}}G)= \si_\rho(F)\star_{\mathbf{m}}\si_\rho(G)\ ,\quad
  \rho\,\si_\rho\circ Q_{\rho^{-1}\mathbf{m}}=Q_{\mathbf{m}}\ ,
  \ee
  we obtain
  \begin{align}\label{Q-zrhoL}
  [Q_{\mathbf{m}},S_{\mathbf{m}}(Z_\rho(iL^\mathbf{m}(g_\eps)))]_{\star_{\mathbf{m}}} & =
  [Q_{\mathbf{m}},\si_\rho\circ S_{\rho^{-1}\mathbf{m}}(iL^{\rho^{-1}\mathbf{m}}(g_{\eps/\rho}))]_{\star_{\mathbf{m}}}\nn\\
  = \rho\,\si_\rho\Bigl([Q_{\rho^{-1}\mathbf{m}},& S_{\rho^{-1}\mathbf{m}}(iL^{\rho^{-1}\mathbf{m}}(g_{\eps/\rho}))
  ]_{\star_{\rho^{-1}\mathbf{m}}}\Bigr)\ .
  \end{align}
By assumption, the adiabatic limit $\eps\downarrow 0$ vanishes for the last expression.
(Due to uniqueness of the adiabatic limit, it does not matter whether we perform this 
limit with $g$ or $g_{1/\rho}$.) With that and with \eqref{Z(L)-adlim} we conclude
$$
  0\approx\lim_{\eps\downarrow 0}\,[Q,S(Z_\rho(iL(g_\eps)))]_\star=
  \lim_{\eps\downarrow 0}\,[Q,S(i\,z_\rho(L)(g_\eps))]_\star =
\bigl[Q,\mathbf{S}[z_\rho(L)]\bigr]_\star\ .\eqno\qedhere
$$
\end{proof}

\paragraph{Perturbative gauge invariance (PGI)}
For the initial model $S(iL(g))$ we admit all renormalization prescriptions which fulfill the 
Epstein-Glaser axioms \cite{EG73,DF04} and 
\textit{perturbative gauge invariance (PGI)} \cite{DHKS94,DS99,Scharf2001,DGBSV10}. 
The latter is a somewhat stronger version of PC, which is formulated \textit{before} the adiabatic 
limit $g\to 1$ is taken. 

In detail, PGI is the condition that to the given interaction $L(g)$ \eqref{L(g)}
  there exists a ``$Q$-vertex''
  \be
 \mathcal{ P}^{\nu}(g;f):=\int dx\,\Bigl(\kappa \,P^\nu_1(x)+\kappa^2g(x)\,P^\nu_2(x)\Bigr) f(x) ,
  \ee
  (where $g,f\in\Dcal(\RR^4)$ and $P_1,\,P_2$ are local field polynomials) 
  and a renormalization of the time-ordered products such that
  \be\label{PGI}
  [Q,S\bigl(i\, L(g)\bigr)]_\star\approx \frac{d}{d\eta}\vert_{\eta =0}\,
  S\bigl(i\, L(g)+\eta\, \mathcal{P}^\nu(g;\d_\nu g)\bigr)\ .
  \ee
The latter equation is understood in the sense of formal power series in $\ka$ and $\hbar$.

That PGI implies PC, is easy to see (on the heuristic level on which we treat 
the adiabatic limit in this paper): the r.h.s.~of \eqref{PGI} vanishes in the adiabatic limit,
since it is linear
in the $Q$-vertex, the latter is linear in $\d_\nu g$ and $\d_\nu g_\eps\sim\eps$.    

Requiring PGI, renormalizability and some obvious properties as Poincar{\'e} invariance and 
relevant discrete symmetries, the Lagrangian of the Standard model of electroweak interactions
has been derived in \cite{DS99,ADS99}. In this way the 
presence of Higgs particles and chirality of fermionic interactions
can be understood without recourse to any geometrical or
group theoretical concepts (see also \cite{StoraVienna1997}).

It is well-known that the $U(1)$-Higgs model is anomaly-free. 
Hence, our initial model can be renormalized such that PGI \eqref{PGI} holds true
for all values of $m,m_H>0$. Using Theorem \ref{thm:PC}, we conclude 
that this model is consistent at all scales.

\section{Higgs mechanism at all scales to 1-loop order}\label{sec:1-loop}

In this section we explain, how one can fulfill the validity of the Higgs mechanism at all scales,
i.e.~the equations \eqref{e-geom-interpret1}-\eqref{e-geom-interpret12}, on 1-loop level.
%For this purpose we derive a lot of results about the 1-loop coefficients $e_\rho^{(1)}$ \eqref{powerseries}
%of the running interaction $z_\rho(L)$ \eqref{z(L)}. Throughout we choose Feynman gauge 
%$\La =1$ for the initial $U(1)$-Higgs model. The conventions for the signs and factors $i,\,2\pi$ 
%are fixed in \eqref{conventions}.

\subsection{The two ways to renormalize}
\label{ssec:explicit-comput}

To write the fundamental formula \eqref{main-theorem} to $n$-th order, we use the chain rule:
\begin{align}\label{eq:Zn}
&Z_{\rho,\mathbf{m}}^{(n)}\bigl(L(g)^{\otimes n}\bigr)=\sigma_\rho \circ T_{n\,\mathbf{m}/\rho}\bigl(
(\sigma_\rho^{-1}\,L(g))^{\otimes n}\bigr)-T_{n\,\mathbf{m}}\bigl(L(g)^{\otimes n}\bigr)\notag\\
&\quad\quad -\sum_{P\in\mathrm{Part}(\{1,...,n\},\,n>|P|>1} T_{|P|\,\mathbf{m}}\bigl(\otimes_{I\in P}
Z_{\rho,\mathbf{m}}^{|I|}(L(g)^{\otimes|I|})\bigr)\ ,
\end{align}
where $Z_\rho^{(n)}:=Z_\rho^{(n)}(0)$ is the $n$-th derivative of $Z_\rho(F)$ at $F=0$ and
the two terms with $|P|=n$ and $|P|=1$, resp., are explicitly written out.

We are now going to investigate the contribution to the r.h.s.~of \eqref{eq:Zn} of a 
primitive divergent diagram $\Ga$, i.e. $\Ga$ has singular order%
\footnote{\label{fn:extension} For $t\in\Dcal'(\RR^l)$ or $t\in\Dcal'(\RR^l\setminus\{0\})$, 
the singular order is defined as $\om(t):=\mathrm{sd}(t)-l$, 
where $\mathrm{sd}(t)$ is Steinmann's scaling degree of $t$, 
which measures the UV-behaviour of $t$ \cite{Ste71}. In the Epstein-Glaser framework, 
renormalization is the extension of a distribution
$t^\circ\in\mathcal{D}'(\RR^l\setminus\{0\})$ to a distribution $t\in
\mathcal{D}'(\RR^l)$, with the condition that $\mathrm{sd}(t)=\mathrm{sd}(t^{\circ})$.
In the case $\mathrm{sd}(t^{\circ})<l$,
the extension is unique, due to the scaling degree requirement, and obtained by
``direct extension'', see \cite[Theorem 5.2]{BF00}, \cite[Appendix B]{DF04} and 
\cite[Theorem 4.1]{DFKR14}.} 
$\omega(\Ga)\geq 0$ and does not contain any subdiagram $\Ga_1\subset\Ga$ with 
less vertices and with $\omega(\Ga_1)\geq 0$. For such a diagram, the expression in 
the second line of \eqref{eq:Zn} vanishes.

Denoting the contribution of $\Ga$ to $T_{n\,\mathbf{m}}\bigl(L(g)^{\otimes n}\bigr)$ by
$$
\int dx_1\ldots dx_n\,\,t_\mathbf{m}^\Ga(x_1-x_n,\ldots,x_{n-1}-x_n)\,P^\Ga(x_1,\ldots,x_n)
\,\prod_{k=1}^n(\ka g(x_k))^{j_k}\ 
$$
(where $P^\Ga(x_1,\ldots,x_n)$ is a, in general non-local, field monomial and the values of
$j_1,\ldots,j_n\in\{1,2\}$ depend on $\Ga$), the computation of the contribution of $\Ga$ 
to $Z_{\rho,\mathbf{m}}^{(n)}\bigl(L(g)^{\otimes n}\bigr)$ amounts to the computation of
\be\label{Ga-scaling}
\rho^{D^\Ga}\,t_{\mathbf{m}/\rho}^\Ga(\rho y)-t_\mathbf{m}^\Ga(y)\ ,
\ee
where $D^\Ga:=\om(\Ga)+4(n-1)\in\NN$ and $y:=(x_1-x_n,\ldots,x_{n-1}-x_n)$.

For simplicity we assume that $0\leq\om(\Ga)<2$; this assumption is satisfied for 
all 1-loop calculations which are done in \cite{Duetsch15} and whose results are used
in this paper.
Applying the scaling and mass expansion (``sm-expansion'')
\cite{Duetsch14}, we then know that $t_\mathbf{m}^\Ga$ is of the form
\be\label{Ga-sm}
t_\mathbf{m}^\Ga(y)=t^\Ga(y)+r_\mathbf{m}^\Ga(y)\ ,\quad r_\mathbf{m}^\Ga=\mathcal{O}(\mathbf{m}^2)\ ,
\quad \om(r_\mathbf{m}^\Ga)<0\ ,
\ee
where $t^\Ga:=t_{\mathbf{m}=\mathbf{0}}^\Ga$ (i.e.~all Feynman propagators are replaced by their 
massless version). The remainder scales homogeneously, 
$\rho^{D^\Ga}\,r_{\mathbf{m}/\rho}^\Ga(\rho y)=r_\mathbf{m}^\Ga(y)$, because it can be renormalized by 
direct extension (see footnote \ref{fn:extension}).

To investigate $\rho^{D^\Ga}\,t^\Ga(\rho y)-t^\Ga(y)$, we omit the upper index $\Ga$ 
and use the notations $\om:=\om(\Ga)$, $l:=(n-1)$ and $Y_j:=y_j^2-i0$. We start with the 
unrenormalized version $t^\circ\in\Dcal'(\RR^{4l}\setminus\{0\})$ of $t:=t^\Ga$, which
scales homogeneously:
\be
\rho^{\om+4l}\,t^\circ(\rho y)=t^\circ(y)\ .
\ee
We work with an analytic regularization \cite{Hollands08}:
\be
t^{\zeta\circ}(y):=t^\circ(y)\,(M^{2l}Y_1\ldots Y_l)^\zeta\ ,
\ee
where $\zeta\in\CC\setminus\{0\}$ with $|\zeta|$ sufficiently small,
and $M>0$ is a renormalization mass scale. $t^{\zeta\circ}$ scales also 
homogeneously -- by the regularization we gain that the degree (of the scaling)  
is $(\om+4l-2l\zeta)$, which is not an integer. Therefore,
the homogeneous extension $t^\zeta \in\Dcal'(\RR^{4l})$ is unique and can explicitly be written
down by differential renormalization \cite[Sect.~IV.D]{DFKR14}.

Using minimal subtraction for the limit $\zeta\to 0$ we obtain an admissible extension 
$t^M \in\Dcal'(\RR^{4l})$ of $t^\circ$ \cite[Corollary 4.4]{DFKR14}:
\begin{align}\label{t-M}
t^M(y)=\frac{(-1)^\om}{\om!}&\sum_{r_1\ldots r_{\om+1}}
\d_{y_{r_{\om+1}}}\ldots\d_{y_{r_1}}\Bigl[\frac{1}{2l}
\Bigl(\overline{y_{r_1}\ldots y_{r_{\om+1}}\,t^\circ(y)\,\log(M^{2l}Y_1\ldots Y_l)}\Bigr)\notag\\
&+(\sum_{j=1}^\om\frac{1}{j})\Bigl(\overline{y_{r_1}\ldots y_{r_{\om+1}}\,t^\circ(y)}\Bigr)\Bigr]\ ,
\end{align}
where $\sum_r \d_{y_r}(y_r\ldots):=\sum_r \d^{y_r}_\mu(y_r^\mu\ldots)$ and the overline denotes the direct 
extension. By means of \eqref{Ga-sm} we obtain the corresponding distribution of 
the massive model: $t_\mathbf{m}^M:=t^M(y)+r_\mathbf{m}$. In the following we use that
$$
\frac{(-1)^\om}{\om!}\sum_{r_1\ldots r_{\om+1}}\d_{y_{r_{\om+1}}}\ldots\d_{y_{r_1}}
\Bigl(\overline{y_{r_1}\ldots y_{r_{\om+1}}\,t^\circ(y)}\Bigr)
=\sum_{|a|=\om}C_a\,\d^a\delta(y)
$$
for some $M$-independent numbers $C_a\in\CC$, as explained after formula (104) in \cite{DFKR14}.

Whether the expression \eqref{Ga-scaling} vanishes depends on the following choice:
\begin{itemize}
\item[(A)] if we choose for $M$ a fixed mass scale, which is independent of $m,m_H$, 
homogeneous scaling is broken:
\be\label{diffren-scal}
\rho^{\om+4l}\,t_{\mathbf{m}/\rho}^M(\rho y)-t_\mathbf{m}^M(y)=
\rho^{\om+4l}\,t^M(\rho y)-t^M(y)=\log\rho\,\,\sum_{|a|=\om}C_a\,\d^a\delta(y)\ ,
\ee
The breaking term is unique, i.e.~independent of $M$; 
therefore, we may admit different values
of $M$ for different diagrams, however, all $M$'s must be independent of $m,m_H$.

\item[(B)] In contrast, choosing $M$ such that it is subject to our scaling transformation, i.e. 
$M:=\al_1 m+\al_2 m_H$ where $(\al_1,\al_2)\in(\RR^2\setminus\{(0,0)\})$
may be functions of $\tfrac{m}{m_H}$, the diagram $\Ga$ does not contribute to the RG-flow:
\be\label{diffren-scal-1}
\rho^{\om+4l}\,t_{\mathbf{m}/\rho}^{M/\rho}(\rho y)-t_\mathbf{m}^M(y)=
\rho^{\om+4l}\,t^{M/\rho}(\rho y)-t^M(y)=0\ .
\ee
\end{itemize}

\begin{rem}\label{AB-PGI} 
The requirement that the initial $U(1)$-Higgs model fulfills PGI,
is neither in conflict with method (A) nor with method (B), for the following 
reason: we require PGI only for the initial model.  
Now, working at one fixed scale, the renormalization constant $M$ 
appearing in \eqref{t-M} may have any value $M>0$ for both methods (A) 
and (B) and, hence, one may choose it such that PGI is satisfied. These methods only prescribe
how $M$ behaves under a scaling transformation: using (A) it remains unchanged, using (B) it 
is also scaled: $M\mapsto\rho^{-1}M$.
\end{rem}

\subsection{Equality of certain coefficients to 1-loop order}
\label{ssec:e=e}

We explain the basic idea in terms of the two diagrams
\begin{align*}
t_{1\,\mathbf{m}}^\circ(y):=&\omega_0\Bigl(T_2(A^\mu\vf(x_1)\otimes A^\nu\vf(x_2))\Bigr)=-\hbar^2g^{\mu\nu}\,
\Delta^F_m(y)\,\Delta^F_{m_H}(y)\ ,\\
t_{2\,\mathbf{m}}^\circ(y):=&\omega_0\Bigl(T_2(A^\mu B(x_1)\otimes A^\nu B(x_2))\Bigr)=-\hbar^2g^{\mu\nu}\,
(\Delta^F_m(y))^2\ ,
\end{align*}
$t_{1\,\mathbf{m}}^\circ , t_{2\,\mathbf{m}}^\circ\in\Dcal^\prime(\RR^{4}\setminus\{0\})$, where
$\om_0$ denotes the vacuum state and $y:=x_1-x_2$. 
These diagrams are related by the exchange of an inner $\vf$-line
with an inner $B$-line. The essential point is that in the sm-expansion of these two distributions,
\be\label{sm-1-2}
t_{j\,\mathbf{m}}^\circ(y)=t_{j}^\circ(y)+r_{j\,\mathbf{m}}^\circ(y)\ ,
\quad r_{j\,\mathbf{m}}^\circ =\mathcal{O}(\mathbf{m}^2)\ , 
\quad \om(r_{j\,\mathbf{m}}^\circ)<0\ ,\quad j=1,2\ ,
\ee
the first term (which is the corresponding massless distribution) is the same: 
$t_{1}^\circ(y)=(D^F(y))^2=t_{2}^\circ(y)$. 

Renormalization is done by extending each term on the r.h.s.~of \eqref{sm-1-2} individually and 
by composing these extensions: $t_{j\,\mathbf{m}}:=t_{j}+r_{j\,\mathbf{m}}\in\Dcal^\prime(\RR^{4})$. 
For the remainders $r_{j\,\mathbf{m}}^\circ$ the direct extension applies (see footnote \ref{fn:extension}),
which maintains homogeneous scaling: $\rho^4\,r_{j\,\mathbf{m}/\rho}(\rho y)=r_{j\,\mathbf{m}}(y)$.
We conclude: if we renormalize $t_{1}^\circ$ and $t_{2}^\circ$ both by  method (A) or 
both by method (B), we obtain
\begin{align*}
\rho^4\,t_{1\,\mathbf{m}/\rho}(\rho y)-t_{1\,\mathbf{m}}(y)&=\rho^4\,t_{1}(\rho y)-t_{1}(y)\\
&=\rho^4\,t_{2}(\rho y)-t_{2}(y)=\rho^4\,t_{2\,\mathbf{m}/\rho}(\rho y)-t_{2\,\mathbf{m}}(y)\ .
\end{align*}
We point out that different renormalization mass scales $M$ for $t_{1}$ and $t_{2}$
are admitted, only their behaviour under the scaling transformation must be the same.
Therefore, this renormalization prescription is compatible with PGI of the initial $U(1)$-Higgs model.

Renormalizing certain Feynman diagrams, which go over into each other
by exchanging $B\leftrightarrow\vf$ for some lines, by the same method (in this sense) --
also triangle and square diagrams with derivatives are concerned -- we obtain that some of the coefficients
$e_\rho$ agree to 1-loop order:
\be\label{e-geom-simplified}
c^{(1)}_{0\rho}=b^{(1)}_{0\rho}\ ,\,\,\, l^{(1)}_{1\rho}=l^{(1)}_{2\rho}\ ,\,\,\,
l^{(1)}_{3\rho}=l^{(1)}_{4\rho}\ ,\,\,\, l^{(1)}_{5\rho}=l^{(1)}_{6\rho}\ ,\,\,\,
l^{(1)}_{7\rho}=l^{(1)}_{8\rho}=l^{(1)}_{9\rho}\ ,
\ee
for details see \cite{Duetsch15}.
With that the equations \eqref{e-geom-interpret2}, \eqref{e-geom-interpret4}, \eqref{e-geom-interpret6}
and \eqref{e-geom-interpret8}-\eqref{e-geom-interpret9} are fulfilled.

In addition, the condition
\be\label{l11=0}
l^{(1)}_{11\rho}=0\ ,
\ee
which is \eqref{e-geom-interpret11} to 1-loop order, can be derived from 
the stability of PC under the RG-flow, by selecting from \eqref{PC-RG} the local terms
which are $\sim A^2\,A\d u$ and by using results of Appendix A in \cite{DS2000}. 

\subsection{Changing the running interaction by finite renormalization}
\label{ssec:fin-ren}

On our way to fulfil the equations \eqref{e-geom-interpret1}-\eqref{e-geom-interpret12}
on 1-loop level, we may use that the following finite 
renormalizations are admitted by the axioms of causal perturbation theory 
\cite{EG73,DF04,Duetsch14} and that they preserve PGI of the initial model:
to $T_2\bigl(L_1(x_1)\otimes L_1(x_2)\bigr)$ we may add
\begin{align}\label{fin-ren-1-1}
\hbar^2 &\,\Bigl(\al_1\,(\d\vf)^2(x_1)
+\al_2\,m_H^2\,\vf^2(x_1)+\al_3\,F^2(x_1)+\al_4\,(\d A+mB)^2\notag\\
& +\al_5\,\bigl(-m^2\,B^2(x_1)+(\d B)^2(x_1)\bigr)+\al_6\bigl(m^2\,A^2(x_1)-(\d A)^2(x_1)\bigr)
\notag\\
& +\al_7\,m^2\,\bigl(-2\,\ut u(x_1)+A^2(x_1)-B^2(x_1)\bigr)\Bigr)\,\delta(x_1-x_2)\,\log\tfrac{m}{M}\ ,
\end{align}
where $\al_1,\ldots,\al_7\in\CC$ are arbitrary.

These finite renormalizations modify the 
1-loop coefficients $e^{(1)}_\rho$ appearing in $z_\rho(L)$ \eqref{z(L)} as follows:
\begin{align}
a_{0\rho}^{(1)}&\mapsto a_{0\rho}^{(1)}+2i\,\al_3\,\log\rho\ ,\label{fin-re-a0}\\
a_{1\rho}^{(1)}&\mapsto a_{1\rho}^{(1)}-i\,(\al_6+\al_7)\,\log\rho\ ,\label{fin-re-a1}\\
a_{2\rho}^{(1)}&\mapsto a_{2\rho}^{(1)}+i\,(\al_4-\al_6)\,\log\rho\ ,\label{fin-re-a2}\\
b_{0\rho}^{(1)}&\mapsto b_{0\rho}^{(1)}-i\,\al_5\,\log\rho\ ,\label{fin-re-b0}\\
b_{1\rho}^{(1)}&\mapsto b_{1\rho}^{(1)}+i\,(\al_4-\al_5-\al_7)\,\log\rho\ ,\label{fin-re-b1}\\
b_{2\rho}^{(1)}&\mapsto b_{2\rho}^{(1)}+i\,\al_4\,\log\rho\ ,\label{fin-re-b2}\\
c_{0\rho}^{(1)}&\mapsto c_{0\rho}^{(1)}-i\,\al_1\,\log\rho\ ,\label{fin-re-c0}\\
c_{1\rho}^{(1)}&\mapsto c_{1\rho}^{(1)}+i\,\al_2\,\log\rho\ ,\label{fin-re-c1}\\
c_{2\rho}^{(1)}&\mapsto c_{2\rho}^{(1)}-i\,\al_7\,\log\rho\ ,\label{fin-re-c2}
\end{align}
the other coefficients remain unchanged.

 We did not find any further finite renormalizations,
 which fufill, besides the already mentioned conditions, the requirements \\
- that they do not add ``by hand'' novel kind of terms to $(z_\rho(L)-L)$ (see \eqref{z(L)})
as e.g.~terms $\sim\d\ut\d u$ or $\sim m\,\ut u\vf$, and\\
- that the equations  \eqref{e-geom-simplified} are preserved.\\
See \cite{Duetsch15} for details.

\subsection{How to fulfill the Higgs mechanism at all scales}\label{ssec:geom-all-scales}
\label{ssec:geom-interpret}

There are two necessary conditions for the Higgs mechanism at all scales, which 
are crucial, since they cannot be fulfilled by finite renormalizations.

\paragraph{Verification of the first crucial necessary condition}
The condition \eqref{geom-interpret-crucial} reads to 1-loop level
\be\label{geom-interpret-crucial-1}
l_{7\rho}^{(1)}-l_{3\rho}^{(1)}=l_{5\rho}^{(1)}-l_{0\rho}^{(1)}\ .
\ee
Since the admissible finite renormalizations \eqref{fin-ren-1-1} do not modify the coefficients
$l_{j\rho}^{(1)}$, there is no possibility to fulfill \eqref{geom-interpret-crucial-1} in this way. 
However, computing explicitly the relevant coefficients $l_{j\rho}^{(1)}$ by
using the renormalization method (A) for all contributing terms, we find that
\eqref{geom-interpret-crucial-1} holds indeed true. This computation, which is given in \cite{Duetsch15},
involves cancellations of square- and triangle-diagrams -- this
shows that \eqref{geom-interpret-crucial-1} is of a deeper kind than the equalities 
derived in Sect.~\ref{ssec:e=e}.

The identity \eqref{geom-interpret-crucial-1} holds also if certain classes of corresponding diagrams 
are renormalized by method (B).

\paragraph{How to fulfill the second crucial necessary condition} The condition
\eqref{e-geom-interpret12} reads to 1-loop order
\be
b_{2\rho}^{(1)}=\tfrac{1}2 \bigl(a_{2\rho}^{(1)}+b_{1\rho}^{(1)}-a_{1\rho}^{(1)}-b_{0\rho}^{(1)}\bigr)\ .
\label{geom-int-crucial}
\ee
Performing the finite renormalizations \eqref{fin-ren-1-1}, i.e.~inserting  
\eqref{fin-re-a0}-\eqref{fin-re-c2} into \eqref{geom-int-crucial}, we find that all
$\al_j$ drop out -- that is, the condition \eqref{geom-int-crucial} cannot be fulfilled by 
means of these finite renormalizations.

Computing the explicit values for the coefficients
$a_{j\rho}^{(1)},\,b_{j\rho}^{(1)}$ by using method (A) (see \cite{Duetsch15}), we find
that \eqref{geom-int-crucial} does not hold.
Hence, using method (A) throughout, we have $\la_{12\rho}\not= 0$, i.e.~the geometrical interpretation 
\eqref{L-new} is violated by terms $\sim A\d B$.

However, we can fulfill the condition \eqref{geom-int-crucial} by switching the method from (A) to (B)
for all diagrams contributing to $b_{1\rho}^{(1)}$ and a part of the diagrams contributing to $a_{1\rho}^{(1)}$
\cite{Duetsch15}. (This switch concerns also all diagrams 
contributing to $a_{0\rho}^{(1)}$, hence we obtain $a_{0\rho}^{(1)}=0$.)

\paragraph{A family of solutions of the Higgs mechanism at all scales} 
The conditions \eqref{e-geom-interpret1}-\eqref{e-geom-interpret12} can be solved to 1-loop order as 
follows: initially we renormalize all diagrams by using method (A), except for the diagrams just mentioned,
for which we use method (B) to fulfill the second crucial necessary condition \eqref{geom-int-crucial}. 
Then we take into account the possibility to modify the coefficients $e_{\rho}^{(1)}$ by
finite renormalizations \eqref{fin-re-a0}-\eqref{fin-re-c2}. This procedure yields the 
following family of solutions:
\begin{align*}
&a_{0\rho}^{(1)}=2\beta_1\,L_\rho\ ,\quad a_{1\rho}^{(1)}=-4\,L_\rho\ ,\quad
a_{2\rho}^{(1)}=(\beta_2-\beta_3)\,L_\rho\ ,\\
&b_{0\rho}^{(1)}=c_{0\rho}^{(1)}=(2+2l_1)\,L_\rho\ ,\quad
b_{1\rho}^{(1)}=(4+2l_1+\beta_2+\beta_3)\,L_\rho\ ,\\
&b_{2\rho}^{(1)}=(3+\beta_2)\,L_\rho\ ,\quad
c_{1\rho}^{(1)}=-\Bigl(6\frac{m^2}{m_H^2}+5\frac{m_H^2}{m^2}\Bigr)\,L_\rho\ ,
\quad c_{2\rho}^{(1)}=(-1+\beta_3)\,L_\rho\ ,%\notag\\
\end{align*}
\begin{align}\label{e-part-solu}
&l_{0\rho}^{(1)}=-3\,L_\rho\ ,\quad l_{1\rho}^{(1)}=l_{2\rho}^{(1)}=:l_1\,L_\rho\ ,\quad
l_{3\rho}^{(1)}=l_{4\rho}^{(1)}=\Bigl(1-6\frac{m^2}{m_H^2}-5\frac{m_H^2}{m^2}\Bigr)\,L_\rho\ ,\notag\\
&l_{5\rho}^{(1)}=l_{6\rho}^{(1)}=-2\,L_\rho\ ,\quad
l_{7\rho}^{(1)}=l_{8\rho}^{(1)}=l_{9\rho}^{(1)}=\Bigl(2-6\frac{m^2}{m_H^2}-5\frac{m_H^2}{m^2}\Bigr)\,
L_\rho
\end{align}
and $l_{11\rho}^{(1)}=0$,
where $L_\rho:=\tfrac{1}{8\pi^2}\,\log\rho$, the number $l_1$ 
is obtained on computing $l_{1\rho}^{(1)}=:l_1\,L_\rho$
by method (A), and $\beta_1:=i8\pi^2\,\al_3,\,\,\beta_2:=i8\pi^2\,\al_4,\,\,
\beta_3:=i8\pi^2\,\al_6=-i8\pi^2\,\al_7\in\CC$ are parameters with arbitrary values.

The family \eqref{e-part-solu} is by far not the general solution of the conditions
\eqref{e-geom-interpret1}-\eqref{e-geom-interpret12}; in particular, there is the trivial
solution $z_\rho(L)=\tfrac{1}{\hbar}\,(L+\mathcal{O}(\hbar^2))$ (i.e. $e_{\rho}^{(1)}=0\,\,\forall e$),
which is obtained by renormalizing all 1-loop-diagrams by method (B).

To 1-loop order one can even find a non-trivial solution of the clearly stronger property of 
BRST-invariance of $(L_0+z_\rho(L))$; but this requires a very specific combination of the methods 
(A) and (B) for the various 1-loop diagrams and suitable finite renormalizations.
Hence, in general, $s(L_0+z_\rho(L))$ is \textit{not} $\simeq 0$, and also
 $s_0L_0$ is \textit{not} $\simeq 0$; in particular these two statements
hold for the family \eqref{e-part-solu} -- see \cite{Duetsch15}.

\subsection{Frequently used renormalization schemes} 

In conventional momentum space renormalization a frequently used renormalization scheme is
dimensional regularization with minimal subtraction, which preserves BRST-invariance generically. 
% of the time-ordered products in the renormalization process. 
Applied to the 1-loop diagrams of our initial model, this property implies 
that the resulting time-ordered products fulfill PGI.%
\footnote{We are not aware of a proof of this statement, but it is very plausible.
A corresponding statement for higher loop diagrams involves a partial
adiabatic limit, because 
such diagrams contain inner vertices, which are integrated out with $g(x)=1$ in  
conventional momentum space renormalization -- but PGI is formulated \textit{before}
the adiabatic limit $g\to 1$ is taken.} 
Dimensional regularization needs a mass scale $M>0$; which remains in the formulas when 
removing the regularization by using minimal subtraction, and plays the role of the 
renormalization mass scale. Usually $M$ is chosen according to method (A); and the minimal 
subtraction prescription forbids to perform any finite renormalization.
Therefore, using this prescription, the Higgs mechanism is not applicable at an arbitrary 
scale, because the second crucial necessary condition \eqref{geom-int-crucial} is violated.
Relaxing this prescription by admitting the finite PGI-preserving renormalizations
\eqref{fin-re-a0}-\eqref{fin-re-c2}, the violation of \eqref{geom-int-crucial} cannot
be removed.

Another state independent renormalization scheme is the central solution of Epstein and 
Glaser \cite{EG73}. (For 1-loop diagrams this scheme corresponds to BPHZ-subtraction
at $p=0$.) Since the subtraction point $p=0$ is scaling invariant, the central solution 
maintains homogeneous scaling (w.r.t. $(x,\mathbf{m})\to (\rho x,\mathbf{m}/\rho)$;
cf.~\cite[Sec.~2.3]{Duetsch14}); hence, the pertinent RG-flow is trivial.

In the conventional literature one meets also state dependent renormalization 
conditions: e.g.~in the adiabatic limit 
the vacuum expectation values of certain time-ordered products must agree with the
``experimentally'' known values for the masses of stable particles in the 
vacuum, and analogous conditions for parameters of certain vacuum correlation functions.
Since ``experimental'' results are not subject to our scaling transformation,
a lot of diagrams are renormalized by method (A), if we use such a scheme.
To 1-loop level, the validity of the Higgs mechanism at all scales amounts then mainly 
to the question: is it nevertheless possible to fulfill the second crucial necessary 
condition \eqref{geom-int-crucial}, which requires to renormalize certain diagrams
by method (B)? 

\section{Summary and conclusions}

In the Epstein-Glaser framework the obvious way to define the RG-flow is to use the Main Theorem 
in the adiabatic limit \cite{PS82,HW03,DF04,BDF09}: the effect of a scaling transformation 
(scaling with $\rho >0$) can equivalently be expressed by a renormalization of the interaction:
$L\mapsto z_\rho(L)$. The so defined RG-flow $\rho\mapsto z_\rho(L)$ depends on the choice of the 
renormalization mass scale(s) $M>0$ for the various UV-divergent Feynman diagrams: 
if $M$ is subject to our scaling transformation (method (B)) -- e.g.~the mass of 
one of the basic fields -- the pertinent diagram does not contribute to the RG-flow.
In contrast, if $M$ is a fixed mass scale (method (A)), the corresponding diagram yields a unique
(i.e.~$M$-independent), non-vanishing contribution.

Performing the renormalization of the wave functions, masses, gauge-fixing parameter and coupling 
parameters, we obtain a description of the scaled model $L_0+z_\rho(L)$ ($L_0$ denotes the free 
Lagrangian) by a new Lagrangian $L_0^\rho+L^\rho$, which has essentially the same form as the original
one, $L_0+L$; but the basic fields and the parameters are $\rho$-dependent. 
The title of this paper can be reformulated as follows: is the new Lagrangian $L_0^\rho+L^\rho$
derivable by the Higgs mechanism for all $\rho >0$? 

We have investigated this question for the $U(1)$-Higgs model to 1-loop order.
We only admit renormalizations of the initial model which fulfill a suitable form of BRST-invariance
of the time-ordered products -- we work with PGI \eqref{PGI}. The answer depends not only on the
choice of the renormalization method ((A) or (B)) for the various 1-loop Feynman diagrams; the RG-flow can 
also  be modified by finite, PGI-preserving renormalizations \eqref{fin-ren-1-1} of the initial model.
Using this non-uniqueness, we have shown that one can achieve that
the Higgs mechanism is possible at all scales; one can even fulfill
the much stronger condition of BRST-invariance of $L_0+z_\rho(L)$.
But this requires a quite (Higgs mechanism) or very (BRST-invariance) specific
prescription for the choice of the renormalization method ((A) or (B)) for the various 
Feynman diagrams, and for the finite renormalizations. If one uses always method 
(A) -- minimal subtraction is of this kind -- the geometrical interpretation is violated
by terms $\sim A\d B$; weakening this prescription by admitting finite PGI-preserving 
renormalizations, these $A\d B$-terms cannot be removed.

If one accepts the Higgs mechanism as 
a fundamental principle explaining the origin of mass at all scales (although it is 
not understood in a pure QFT framework), our results exclude quite a lot of 
renormalization schemes, in particular minimal subtraction.
%If one wants to keep such a renormalization scheme, one has to call in question the 
%assumptions about the Higgs mechanism or our definition of the RG-flow.

On the other hand we give a model-independent proof, which uses rather
weak assumptions, that the RG-flow is compatible with 
a weak form of BRST-invariance of the time-ordered products, namely PC (Theorem \ref{thm:PC}).
However, in \cite{Duetsch15} it is shown that the 
somewhat stronger property of PGI gets lost under the RG-flow in general,
and in particular if one uses a renormalization prescription corresponding to minimal subtraction.

% ------------------------------------------------------------------------

\subsection*{Acknowledgment}
During working at this paper the author was mainly at the Max Planck Institute 
for Mathematics in the Sciences, Leipzig; he thanks Eberhard Zeidler for the invitations to 
Leipzig and for valuable discussions. In addition, the author profitted from invitations to give
a talk about the topic of this paper at the workshop 
``Algebraic Quantum Field Theory: Its status and its future'' at the Erwin Schr\"oder Institute in Vienna
(19.-23.05.2014) and at the conference ``Quantum Mathematical Physics'' in Regensburg (29.09.-02.10.2014).
The author thanks also the Vicerrector{\'i}a de Investigaci{\'o}n of the 
Universidad de Costa Rica for financial support.
The question in the title of this paper was found during innumerable discussions with
J\"urgen Tolksdorf about his geometrical derivation of a value for the Higgs mass.
The author profitted also a lot from stimulating discussions with Klaus Fredenhagen, 
Jos{\'e} M. Gracia-Bond{\'i}a, Bert Schroer, G\"unter Scharf, Klaus Sibold and Joseph~C. V\'arilly.

% ------------------------------------------------------------------------
\end{document}